# Disorder-broadened phase boundary with enhanced amorphous superconductivity in pressurized In$_2$Te$_5$

Yi Zhao[1#], Tianping Ying[2#*], Lingxiao Zhao[1#], Juefei Wu[1#], Cuiying Pei[1], Jing Chen[2], Jun Deng[2], Qinghua Zhang[2], Lin Gu[3], Qi Wang[1,4], Weizheng Cao[1], Changhua Li[1], Shihao Zhu[1], Mingxin Zhang[1], Na Yu[1], Lili Zhang[5], Yulin Chen[1,4,6], Chui-Zhen Chen[7*], Tongxu Yu[8], and Yanpeng Qi[1,4,9*]

1. School of Physical Science and Technology, ShanghaiTech University, Shanghai 201210, China
2. Institute of Physics and University of Chinese Academy of Sciences, Chinese Academy of Sciences, Beijing 100190, China
3. Beijing National Center for Electron Microscopy and Laboratory of Advanced Materials, Department of Materials Science and Engineering, Tsinghua University, Beijing 100084, China
4. ShanghaiTech Laboratory for Topological Physics, ShanghaiTech University, Shanghai 201210, China
5. Shanghai Synchrotron Radiation Facility, Shanghai Advanced Research Institute, Chinese Academy of Sciences, Shanghai 201203, China
6. Department of Physics, Clarendon Laboratory, University of Oxford, Parks Road, Oxford OX1 3PU, UK
7. Institute for Advanced Study and School of Physical Science and Technology, Soochow University, Suzhou 215006, China
8. Suzhou Laboratory, Suzhou, Jiangsu, 215123, China
9. Shanghai Key Laboratory of High-resolution Electron Microscopy, ShanghaiTech University, Shanghai 201210, China

\# These authors contributed to this work equally.
\* Correspondence should be addressed to Y.Q. (qiyp@shanghaitech.edu.cn) or T.Y. (ying@iphy.ac.cn) or C.C (zchen@suda.edu.cn)

**Abstract**

As an empirical tool in materials science and engineering, the iconic phase diagram owes its robustness and practicality to the topological characteristics rooted in the celebrated Gibbs phase law (F = C − P + 2). When crossing the phase diagram boundary, the structure transition occurs abruptly, bringing about an instantaneous change in physical properties and limited controllability on the boundaries (F = 1). Here, we expand the sharp phase boundary to an amorphous transition region (F = 2) by partially disrupting the long-range translational symmetry, leading to a sequential crystalline-amorphous-crystalline (CAC) transition in a pressurized In$_2$Te$_5$ single crystal. Through detailed *in-situ* synchrotron diffraction, we elucidate that the phase transition stems from the rotation of immobile blocks [In$_2$Te$_2$]$^{2+}$, linked by hinge-like [Te$_3$]$^{2-}$ trimers. Remarkably, within the amorphous region, the amorphous phase demonstrates a notable 25% increase of the superconducting transition temperature ($T_c$), while the carrier concentration remains relatively constant. Furthermore, we propose a theoretical framework revealing that the unconventional boost in amorphous superconductivity might be attributed to an intensified electron correlation, triggered by a disorder-augmented multifractal behavior. These findings underscore the potential of disorder and prompt further exploration of unforeseen phenomena on the phase boundaries.

**Main**

Crystallographic phase boundaries, a concept widely applied in metallurgy, chemistry, and material science, are often underestimated in their importance[1]. This oversight is primarily rooted in the Gibbs phase rule, F (free variables) = C(components) - P (phases) + 2, where the free variable is constrained to 1 (F = 1) on the phase boundaries[2]. Consequently, most crystalline phase transitions entail abrupt changes in both physical and chemical properties across the boundaries (Figure 1a)[3]. However, phase boundaries can be important, such as the well-known "morphotropic phase boundary (MPB)," widely used for enhancing piezoelectricity[4]. Despite ongoing debates about its origin, the application of MPB to other physical properties remains limited. An intriguing, yet unanswered question is whether it is possible to transform an abrupt phase boundary into a broader region through a highly tunable external parameter. An ensuing and more important question is how the corresponding physical properties, such as superconductivity, magnetization, optics, and so on, will change within this broadened region—whether they will increase, decrease, or vary continuously (Figure 1b). According to the Gibbs phase rule, it is natural to suppose that if the crystalline phase transition involves passing through an equilibrium amorphous phase (P = 1) at specific pressure and temperature, the thermodynamic phase boundary can be broadened into a continuous region (F = 2).

Attaining the intermediate amorphous phase poses a formidable challenge. While amorphization of solids and crystallization of liquids under external pressure are commonly observed, the occurrence of a continuous solid-state evolution from a crystalline-amorphous-crystalline (CAC) phase transition is exceptionally rare. $Ge_2Sb_2Te_5$, a phase-change compound, stands as one of the few exceptions, likely due to the existence of rocksalt-like local atomic motifs and randomly oriented Te bonds under pressure[5]. This observation drives our exploration into disrupting long-range translational symmetry while preserving short-range order, enabling the recrystallization of ordered local motifs at higher pressures. Achieving this demands a combination of rigid building blocks that remain relatively unchanged under pressure and more flexible components that can be easily distorted. Our strategy involves leveraging element Te, known for its extended p-orbital that facilitates diverse bond connections. In addition, the covalent bonds within $In_2Te_2$ (Figure 1c) form a parallelogram structure that shared by all known In-Te binary compounds. (Figure S1) Consequently, $In_2Te_5$, with a space group of $Cc$, emerges as a promising candidate, featuring a series of $In_2Te_2$ parallelograms connected by $Te_3$ trimers (Figure 1d, Figure S2).

Here, we present the achievement of a disorder-broadened phase transition in $In_2Te_5$ over a broad pressure range of 10 GPa, marking a sequential CAC phase transition facilitated by the rotation of $In_2Te_2$ blocks. A significant discovery is a bulk response of enhancement of superconducting transition temperature ($T_c$) upon entry into the amorphous phase. Theoretical analysis identifies the crucial role played by the disorder-induced multifractality of the electronic eigenfunctions, attributing it to the marked enhancement in superconductivity. Our results not only highlight the paramount significance of phase boundaries but also emphasize the substantial value in expanding these boundaries, providing unprecedented insights into the uncharted territories of these metastable regions.

**Structure design strategy**

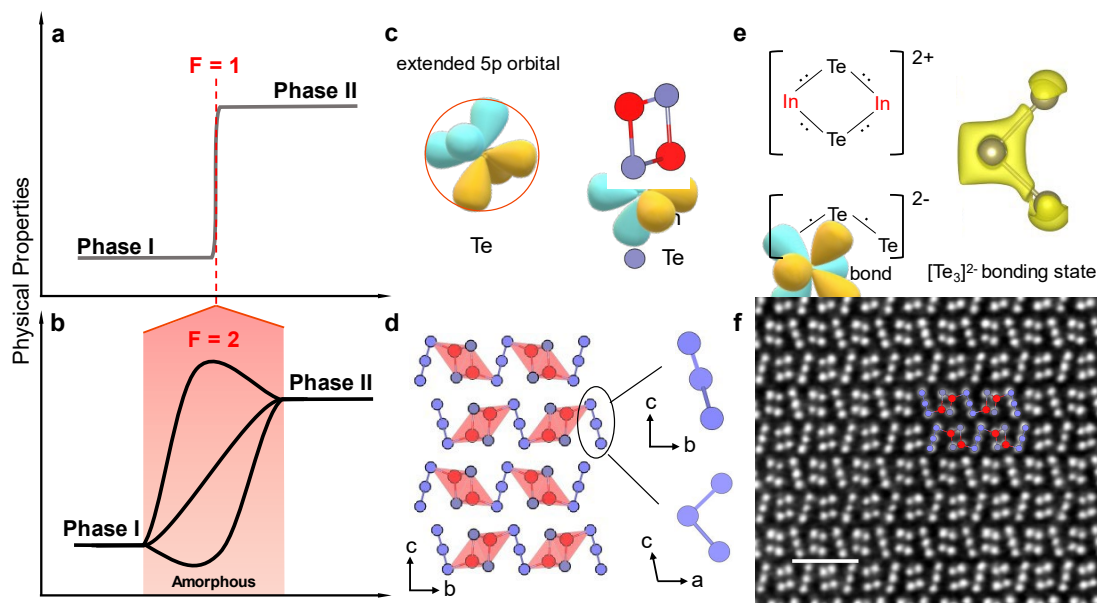

**Figure 1.** Broadened Phase Transition and Structure Design Strategy. **a-b.** Conceptual illustration depicting the broadening of the sharp boundary of a crystalline phase transition into an intermediate region with the variation of an arbitrary tuning parameter, where intrinsic physical properties monotonically evolve or generate local maximum or local minimum. **c.** (left) Illustration demonstrating the spatially extended $5p$ orbital of Te, potentially enabling flexible bond connections, paired with an $In_2Te_2$ parallelogram building block (right) common to all known In-Te binary compounds. **d.** Crystal structure of $In_2Te_5$ viewed along the $a$-axis, with a magnified side-view of the $Te_3$ trimer. **e.** Bonding analysis of $In_2Te_2$ and $Te_3$ trimer. **f.** Atomic-resolved scanning transmission electron microscopy (STEM) micrograph of $In_2Te_5$ along the $a$-axis. The scale bar is 1 nm.

The structure is composed of two distinct subunits: the parallelogram-shaped $In_2Te_2$ and the hinge-like $Te_3$ trimer, illustrated in Figure 1e. Maintaining the standard chemical valence states of $In^{3+}$ and $Te^{2-}$, the $In_2Te_2$ units can be denoted as $[In_2Te_2]^{2+}$, interconnected via edge sharing, resulting in a fully occupied bonding configuration. To maintain charge neutrality, the $Te_3$ trimer incorporates two electrons, forming $[Te_3]^{2-}$. Detailed bonding analysis can be found in Figure S3, and the charge density of the $[Te_3]^{2-}$ is illustrated in Figure 1e, forming a banana-like bonding configuration[6]. This quality grants $[Te_3]^{2-}$ a remarkable degree of flexibility, allowing it to deform easily under mild pressure while maintaining the relatively intact structure of the $[In_2Te_2]^{2+}$ block. To cultivate the $In_2Te_5$ single crystal, we employed the chemical vapor transport (CVT) method as described in Methods. The optical image, X-ray diffraction, and composition analysis of the acquired single crystals are detailed in Figure S4-S8. The atomic-resolved STEM image viewed along the $a$-axis (Figure 1f) verified the atomic structure of $In_2Te_5$ with the space group of $C$c (No. 9).

**Crystalline-amorphous-crystalline phase transition**

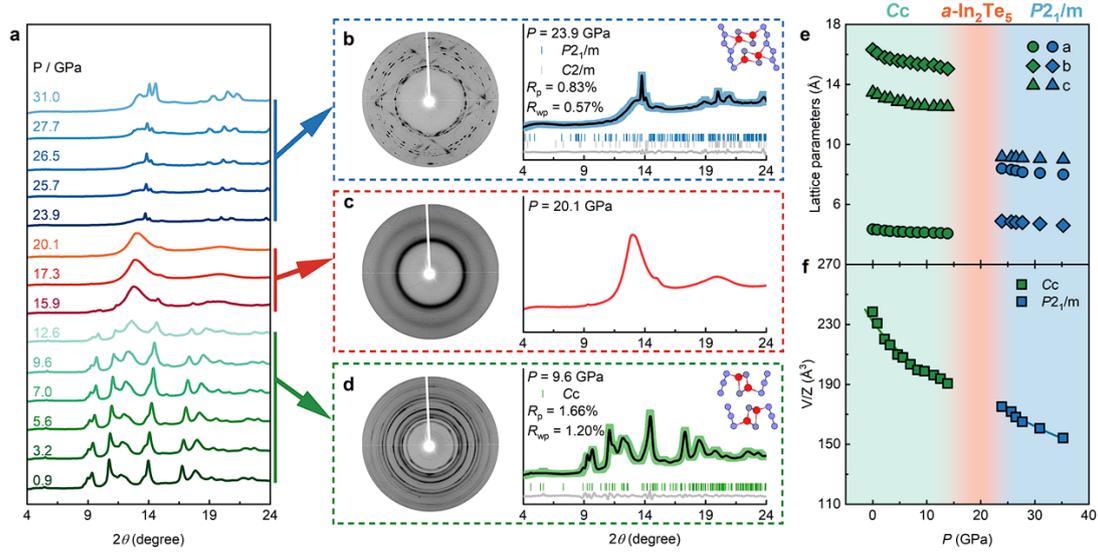

**Figure 2.** Pressure-induced CAC Phase Transition. **a.** *In-situ* synchrotron diffractions with the increase of external pressure. The green, red, and blue-colored regions denote the crystalline, amorphous, and recrystallized regions, respectively. **b-d.** Representative diffraction patterns and their Rietveld refinement at 23.9, 20.1, and 9.6 GPa. The Rietveld refinement results of the high-pressure and low-pressure are shown. Insets of **b** and **d** illustrate the building block of the characteristic structural units. **e,f.** Evolution of lattice constants and cell volume with applied pressure. The red region denotes the amorphous phase, where the diffraction peaks cannot be distinguished.

We investigated the efficacy of the "block-hinge" strategy to achieve a continuous crystalline phase transition by employing *in-situ* synchrotron diffraction while progressively increasing the pressure up to 31 GPa (Figure 2a). Below 12.6 GPa, all diffraction peaks remain preserved but continuously shift to higher angles, indicating the ongoing suppression of the lattice parameter. Beyond 12.6 GPa, the diffraction intensities rapidly diminish, leaving two persistent broad humps at approximately 12° and 21°, which signify the hallmark of pressure-induced amorphization. Notably, upon further increasing the pressure to 23.9 GPa, a series of sharp new peaks emerge from the amorphous background, elucidating the CAC transformation. Representative diffraction patterns in Figure 2b-2d showcase this transformation clearly. Therefore, as the external pressure increases monotonically, the initial $Cc$ structure (Figure 2d at 9.6 GPa) gradually loses its crystallinity, resulting in diffraction patterns that blur into two indistinct rings at 20.1 GPa (Figure 2c). However, at 23.9 GPa, distinct diffraction spots emerge, signifying pressure-induced recrystallization at higher pressures.

For the determination of high-pressure structure, we utilized the established structure search software MACUS based on evolutionary algorithm. A comprehensive analysis of the calculation result is available in Figure S9-S10, which shows the most stable structure is $P2_1/m$ at moderate pressure, which then transits subsequently to a $C2/m$ phase with the increase of pressure. Notably, within a theoretical pressure range of 20-30 GPa, the formation energy of these two phases appears comparable, suggesting the potential coexistence of both phases. This observation aligns with our high-pressure diffraction patterns (Figure 2b), in which the pattern can be effectively fitted using two-phases. Rietveld refinement results indicate that the mole ratio of the $C2/m$ phase is less than

20%. Consequently, it is reasonable to consider the dominant structure beyond $a$-In$_2$Te$_5$ phase is $P2_1/m$, with its functional units depicted in the inset of Figure 2b. Nevertheless, our subsequent measurements suggest that the presence of the minor $C2/m$ phase does not alter our primary conclusion. The pressure-dependent lattice parameters of the $Cc$ phase and $P2_1/m$ phase are depicted in Figure 2e, showcasing a distinct discontinuity across the amorphous In$_2$Te$_5$ ($a$-In$_2$Te$_5$) region. Meanwhile, the calculated cell volume indicates a consistent reduction over the highlighted red area, as illustrated in Figure 2f, validating the correctness of our identified high-pressure structure.

The occurrence of a CAC solid-state transition is quite rare, contrasting the more common instances of pressure-induced amorphization in solid compounds and the solidification of liquids. Only a handful exceptions, such as Ge$_2$Sb$_2$Te$_5$, are known to display such transitions, attributed to the presence of rocksalt-like local motifs and relatively mobile Te-vacancy pairs, which is crucial for achieving an intermediate amorphous state[5c, 7]. Similar phenomenon were also reported in SnI$_2$, although only a portion of the compound turns to a non-crystalline state under pressure[8]. The theoretical prediction or simulation of the structure transformation in the amorphous phase remains a daunting task. Therefore, the successful realization of a CAC-type transition in pressurized In$_2$Te$_5$ stands as a validation of the efficacy of our "block-hinge" strategy.

It is also worth noting that conventional pressure-induced amorphization is typically accompanied by large hysteresis during compression and decompression processes. In such instances, only a unidirectional crystalline-amorphous transition is typically observed, and these phase transitions can no longer be viewed as thermodynamical equilibrium processes. Considering the unique CAC transition observed in pressurized In$_2$Te$_5$, it is imperative to determine whether the Gibbs phase rule still holds. We carried out a detailed Raman measurements encompassing the complete compression-decompression cycle, as shown in Figure S11. Interestingly, we discovered that the crystalline state is recovered during the decompression process with negligible pressure hysteresis. This may stem from the retention of relatively rigid In$_2$Te$_2$ units within the amorphous phase, a characteristic distinct from other pressure-induced amorphous phases in which all atoms are arranged randomly.

# Microscopic understanding of structure evolution

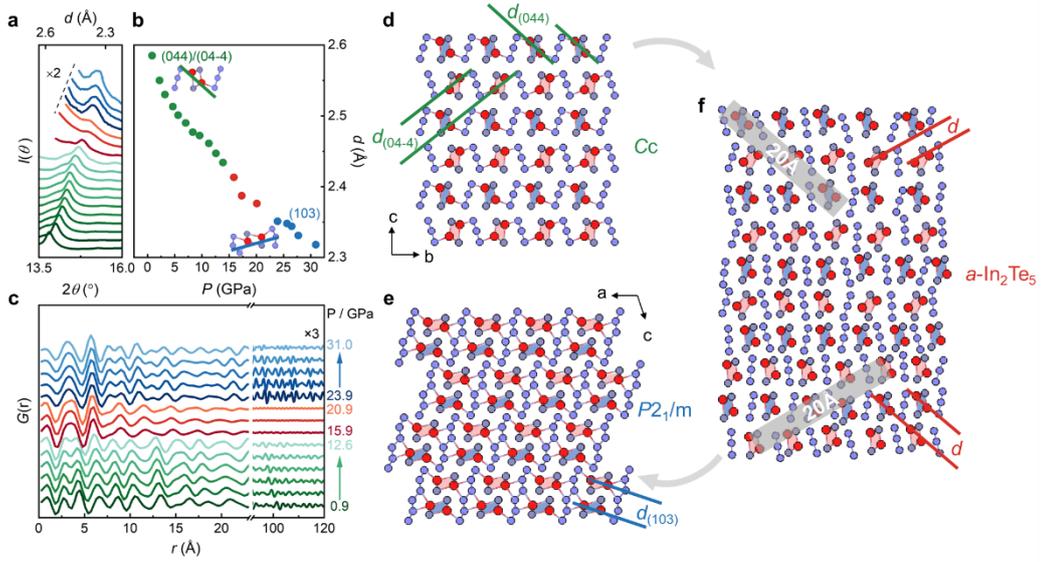

**Figure 3.** Microscopic understanding of structure evolution with preserved local structural ordering. **a.** A zoomed-in view of the synchrotron diffraction patterns shown in Figure 2a in the range of 13.5°-16°. A distinguishable peak can be observed even in the amorphous phase. This peak corresponds to the (044)/(04-4) crystal plane in the $Cc$ phase and (103) plane in the $P2_1/m$ phase. **b.** The character distance shows a continuous contraction with the applied pressure which corresponds to the interplanar distance of the In-In plane shown in the inset. **c.** Pair distribution functions (PDF) with the increasing of external pressure. The broken horizontal axis omits the featureless region of the amorphous phase, beyond which clear peaks in the $Cc$ and $P2_1/m$ phases can still be seen. **d-f.** Illustration of the microscopic structure evolution with irregular rotational $In_2Te_2$ blocks, thanks to the flexible $Te_3$ banana bond connections. The formation of $a$-$In_2Te_5$ phase originates from the loss of long-range translational symmetry.

We next scrutinize the CAC transformation in pressurized $In_2Te_5$. If amorphization in $In_2Te_5$ occurs at the cluster level rather than the atomic scale, certain characteristic length features should be observable. Hence, we revisited the synchrotron diffraction patterns and noticed a small yet discernible peak in the $a$-$In_2Te_5$ phase in Figure 2a. Figure 3a provides a zoomed-in view of the diffraction patterns in the range of 13.5° to 16°, where the peak remains across all three phases. These peaks correspond to the (044) and (04-4) faces of the $Cc$ phase and the (103) faces of the $P2_1/m$ phase, primarily containing the interplanar distance of the In-In plane (Figure 3b). The presence of these peaks in the $a$-$In_2Te_5$ phase indicates the existence of short-range ordering.

As shown in Figure 3c, the high-pressure *in-situ* pair distribution functions (PDF) show huge differences around the phase transition pressures. In the pressure range from 0.9 to 12.6 GPa, the two peaks observed at 2.5 and 4.5 Å are gradually merging, and they suddenly turn into one broad peak after transforming into the amorphous state. Upon subsequent transformation to the $P2_1/m$ phase, some new peaks emerge. Beyond 20 Å, the PDFs of the $Cc$ phase and the $P2_1/m$ phase still show clear peaks, which is a typical crystalline behavior. However, in the amorphous phase, diffraction peaks quickly dissipate, indicating the absence of long-range order. In contrast, clear peaks can be observed within 20 Å.

Interestingly, the series of peaks even exist in the amorphous phase, which gives evidence for the existence of local structures. Considering the structures of the $Cc$ phase and the $P2_1/m$ phase shown in Figure 3d and e, the main difference is the orientation of the $In_2Te_2$ parallelograms. To be more specific, the In-In bonding directions of the adjacent layers are inclined at an angle in the $Cc$ phase, but all of them are parallel in the $P2_1/m$ phase. As illustrated in Figure S14-S17, the rotation of inclined crystal planes can readily transform the $Cc$ phase to $P2_1/m$. Owing to the flexibility of the $Te_3$ trimer, this transition naturally takes place in a random manner. The characteristic length of 20 Å covers 3-4 $In_2Te_2$ parallelograms in the $a$-$In_2Te_5$ phase, a typical length scale of amorphous materials. This observation suggests that the local structural motifs of $a$-$In_2Te_5$ are not completely disordered, rather, there are parallel In-In directions within each layer at nanometer scale. This 'hinge-block' structure effectively preserves the local structures of $In_2Te_2$ parallelograms during the recrystallization from $a$-$In_2Te_5$ to the $P2_1/m$ phase. Therefore, the structural evolution of $Cc$-amorphous-$P2_1/m$ can be summarized in two steps. Initially, the arbitrary distortion of the $Te_3$ trimer leads to the amorphization of the $Cc$ phase into $a$-$In_2Te_5$. Subsequently, the relatively robust $In_2Te_2$ parallelograms restack in a compact manner at higher pressures and recrystallize into the $P2_1/m$ phase. An illustration of the amorphous structure is shown in Figure 3f.

**Disorder-enhanced superconductivity**

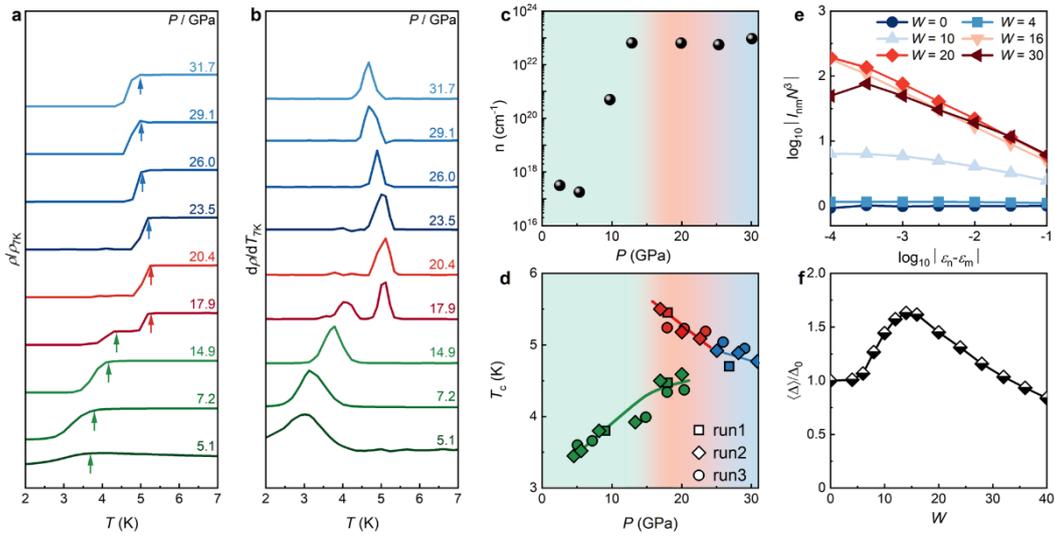

**Figure 4.** Disorder-enhanced superconductivity in the broadened phase transition region. **a-b.** Normalized resistivity and their derivatives as pressure varies. Arrows mark the onset $T_c$. **c, d.** Carrier concentration and extracted $T_c$ versus the increasing of pressure. **e-f.** Theoretical simulations of the disorder-dependent superconducting gaps $\langle \Delta \rangle$ and eigenstates correlations $I_{nm}$. The correlations increase when the eigenstates overlap in a more confined space, amplifying the effective interaction and leading to a larger superconducting gap. Here, $\langle \Delta \rangle$ and $\Delta_0$ denote the spatial-averaged superconducting gap with and without disorder, respectively.

As we successfully broaden the sharp phase boundary into a continuous phase region, the accompanying properties will not change abruptly across the amorphous region. Therefore, it is tempting to trace the influence of numerous physical properties, among which electronic transport measurements are relatively easy to perform considering their feasibility under pressure. As shown in Figure 4a and b, superconductivity emerges at 5.1 GPa, with a $T_c$ of 3.6 K in the $Cc$ phase. The observation of an insulating-superconducting transition is unsurprising, given the existence of many telluride superconductors[9]. Interestingly, a kink in resistance at a much higher temperature of 5.5 K emerges at 17.9 GPa, the pressure at which $In_2Te_5$ starts to transform into the amorphous phase, which is followed by a second resistance step at 4.5 K. A detailed characterization of the higher $T_c$ can be found in Figure S18-S28, while the lower $T_c$ corresponds to the remaining $Cc$ phase, aligning well with the residual $Cc$ phase observed in the XRD pattern shown in Figure 2a at 15.9 GPa. With a further increase in pressure, e.g., at 20.4 GPa, only one transition can be observed, consistent with the completion of amorphization under pressure. At higher pressure, the $T_c$ monotonically decreases in the $P2_1/m$ phase. We further measure the variation of the carrier concentration with the applied external pressure. As shown in Figure 4c, the carrier density in the $Cc$ phase quickly increases with applied pressure and saturates at 12.9 GPa, corresponding well with the gradual increase in $T_c$ and the subsequent saturation upon entering the amorphous region (green curve). This implies that the observed insulating-superconducting transition and the subsequent enhancement of $T_c$ in the $Cc$ phase are caused by pressure-induced band overlapping and the subsequent increase in carrier density. On the contrary, the emergence of a 25% increase in $T_c$ in the $a$-$In_2Te_5$ phase defies attribution to the increase of electron concentration, which saturates prior entering the highly disordered region (Figure 4d).

The interplay between disorder and superconductivity is nuanced and intriguing, given the general tendency of electrons to be confined in disordered systems [10], while superconductors typically facilitate unimpeded transmission of electrons. It is rare to observe enhanced superconductivity in an amorphous system, but notable exceptions exist, such as bismuth, which exhibits a relatively high $T_c$ of 6 K in its amorphous form compared to a $T_c$ below 0.5 mK in crystals [11]. The underlying mechanism remains unclear. Given the propensity for amorphous bismuth to readily crystallize above 12 K [12], the stable $a$-$In_2Te_5$ phase offers an ideal platform to untangle this enigma. Another illustrative example is the surface-doped $NbSe_2$ monolayer, where, after eliminating various mechanisms such as carrier doping and suppression of CDW ordering, the substantial increase in superconductivity is attributed to the multifractality of the eigenfunction[13]. Meanwhile, the creation of a broadened phase transition region does not guarantee an enhanced $T_c$. For instance, $Ge_2Sb_2Te_5$, which also undergoes a CAC transition and shows superconductivity under high pressure. In this case, its $T_c$ smoothly evolves across the amorphous region, reflecting the monotonic progression of physical properties depicted in Figure 1b[7c].

To understand the enhancement of superconductivity arising from disorder, we investigate the interplay between superconductivity and inhomogeneities by conducting precise self-consistent calculations of the modified Bardeen-Cooper-Schrieffer (BCS) mean-field Hamiltonian. Detail of our model Hamiltonian and self-consistent gap equation can be found in the Supplementary Information. The zero-temperature self-consistent gap equation is given by

$$\Delta_m = \frac{U}{2} \sum_{|\varepsilon_n|<\varepsilon_D} \frac{1}{\sqrt{\varepsilon_n^2 + \Delta_n^2}} I_{nm}\Delta_n$$

where $\Delta_m$ denotes the superconducting paring potential, $U$ signifies the intensity of the attractive interaction, and $\varepsilon_D$ stands for the Debye energy. The interaction matrix $I_{nm} = \sum_i \phi_{m\uparrow}^*(r_i)\phi_{\bar{m}\downarrow}^*(r_i)\phi_{n\uparrow}(r_i)\phi_{\bar{n}\downarrow}(r_i)$ describes eigenstates correlations at different energies, and $\phi_{n\uparrow}(r_i)$ represents eigenstates of the non-interacting Hamiltonian associated with the eigenvalue $\varepsilon_n$. Typically, $I_{nm}$ increases when eigenfunctions overlap in a more confined space, amplifying the effective interaction and leading to a larger $\Delta_m$. As shown in Figure 4e, in the low disorder regime ($W < 4$), $I_{nm} \sim 1/N^3$ is independent of disorder, because the eigenstate is approximately extended. This corresponds to the regime in which Anderson's theorem applies [14], i.e. the spatial averaged paring potential $\langle\Delta\rangle$ is independent of disorder strength (see the $W < 4$ regime in Figure 4f). However, when the disordered system approaches the metal-insulator transition $W \approx 16$, the eigenfunctions are multifractal[15]. In the case, the matrix $I_{nm}$ can be approximated by $I_{nm} = \frac{1}{N^3}\left|\frac{E_0}{\varepsilon_n - \varepsilon_m}\right|^\gamma$ for a $N \times N \times N$ cubic system (see Figure 4e), when $\delta_0 < |\varepsilon_n - \varepsilon_m| < E_0$ with $\delta_0$ the mean level space and $E_0$ the energy cutoff of fractal behaviors. Therefore, $I_{nm}$ and thus $\Delta_m$ increases with $\gamma$, where $\gamma = 1 - \frac{d_q}{d} > 0$ with the fractal dimension $d_q < d = 3$. As a result, this enhancement of $I_{nm}$ due to multifractality of the eigenfunction can lead to an increase $\Delta_m$ for $4 < W < 16$ as shown in Figure 4f. In the experiment, the electrons exhibit growing multifractality as the system transitions towards the increasingly disordered amorphous regime. This gives rise to the unconventional boost in superconductivity in the amorphous regime near 20 GPa shown in Figure 4d.

**Conclusion**

In summary, our study effectively leverages the concept of partial disorder to introduce an intermediate amorphous phase, leading to a substantial enhancement of superconductivity during the transition between two consecutive crystalline solid phases in pressurized In$_2$Te$_5$ single crystal. Employing *in-situ* synchrotron diffraction and PDF analysis, we elucidate the structural evolution involving the rotation of rigid In$_2$Te$_2$ parallelograms connected by flexible Te$_3$ trimers. A qualitative understanding of the enhanced $T_c$ is attained through a multifractal perspective, mirroring electron wave vector overlap observed in strongly correlated materials. Importantly, Figs. 1a-b transcend the realm of superconductivity, as the broadening of the phase boundary may hold implications across various branches, including magnetism, ferroelectrics, optical properties, and more. The proposed "hinge-block" strategy not only offers a promising approach for designing candidates undergoing the CAC transition but also unveils avenues for exploring the concealed intricacies of boundary phenomena.

## Methods

### Crystal growth

High-quality single crystals of $In_2Te_5$ were synthesized by chemical vapor transport method[16]. The acquired single $In_2Te_5$ crystals were needle shaped. A suitable crystal was selected and analyzed on a Bruker APEX-II CCD diffractometer. The crystal was kept at 150 K during data collection. Using Olex2[17], the structure was solved with the SHELXT[18] structure solution program using Intrinsic Phasing and refined with the SHELXL[19] refinement package using Least Squares minimization. Scanning electron microscope images were acquired using the JSM-IT500HR/LA, with an acceleration voltage of 10 keV and a work distance of 11 mm. Scanning transmission electron microscopy (STEM) measurements are carried out on $In_2Te_5$. The atomic structures of the $In_2Te_5$ was characterized using an ARM−200CF (JEOL, Tokyo, Japan) transmission electron microscope operated at 200 kV.

### High-pressure measurement

High-pressure in-situ electrical transport property was performed in Physical Property Measurement System (PPMS-9T). Nonmagnetic diamond anvil cell (DAC) was employed to perform the *in situ* high-pressure resistivity measurements. A cubic BN/epoxy mixture was used as insulating layer between BeCu gaskets and electrical leads. Four Pt foils were arranged in a van der Pauw four-probe configuration to contact the sample in the chamber for resistivity measurements[20]. Pressure was determined by the ruby luminescence method[21]. The *in situ* high-pressure Raman spectroscopy measurements were performed using a Raman spectrometer (Renishaw inVia, U.K.) with a laser excitation wavelength of 532 nm and low-wavenumber filter. Symmetric DAC with anvil culet sizes of 300 μm was used, with silicon oil as pressure transmitting medium (PTM)[22]. *In situ* high-pressure synchrotron x-ray diffraction (XRD) measurements were carried out at room temperature with sample powder grinded from single crystals at the beamline BL15U of Shanghai Synchrotron Radiation Facility (X-ray wavelength $\lambda = 0.6199$ Å). Symmetric DACs with anvil culet sizes of 300 μm and T301 gaskets were used. Daphne 7373 oil was used as the PTM and pressure was determined by the ruby luminescence method[21]. The two-dimensional diffraction images were analyzed using the Dioptas software[23]. Rietveld refinements on crystal structures under high pressure were performed using the General Structure Analysis System (GSAS) and the graphical user interface EXPGUI[24].

### Structure search

We utilized the established structure search software MACUS[25] to determine the high-pressure structures of $In_2Te_5$, with their respective thermodynamic stabilities analyzed. We performed the crystal structure searches under 20GPa and 50GPa, each evolution was implemented for 25 generations with 30 structures per generation. The cutoff energy of the plane-wave was set to 350 eV and the sampling grid spacing of the Brillouin zone was $2\pi \times 0.05$ Å$^{-1}$ in structure searching. The structure optimization and electronic structure calculations were carried out by the Vienna Ab initio Simulation Package (VASP) based on the density functional theory[26]. The exchange-correlation functional is treated by the generalized gradient approximation of Perdew, Burkey, and Ernzerhof[27]. The calculations use projector-augmented wave (PAW) [28] approach to describe the core electrons and their effects on valence orbitals. To describe the van der Waals (vdW) interactions, we used the rev-vdW-DF2 (also known as vdW-DF2-B86R) [29] function. The spin-orbit coupling

(SOC) is also taken into account. We set the plane-wave kinetic-energy cutoff to 450 eV, and the Brillouin zone was sampled by the Monkhorst-Pack scheme of $2\pi \times 0.03$ Å$^{-1}$. The convergence tolerance was $10^{-6}$ eV for total energy and 0.003 eV/Å for all forces. Phonon spectrum calculations were performed by utilizing the supercell finite displacement method implemented in the PHONOPY [30] package, and $2 \times 2 \times 2$ supercell was applied for the predicted structures.

**PDF analysis**

The pair distribution functions (PDF) are extracted from the high-pressure XRD results, by using the PDFgetX3 software[31]. The background signals of the XRD results have already been subtracted. Since the resolution and range of the diffraction images are 0.003 Å$^{-1}$ and 0.55~4.3 Å$^{-1}$, respectively, the interval and range of the acquired PDFs are chosen to be 0.3 Å and 0 – 500 Å.

**The model Hamiltonian and self-consistent gap equation**

The negative-U Hubbard Hamiltonian (or the space-dependent mean-field BCS Hamiltonian) is a widely used Hamiltonian to study the interplay of superconductivity and inhomogeneities[32]. The Hamiltonian mainly includes three parts: (i) the electron hopping terms, which describe the free electrons on the Fermi surfaces; (ii) the negative-U Hubbard term, which denotes the electron-electron attractive interaction; and (iii) the random potential, which accounts for the impurities in dirty superconductors. It is noted that the attractive interaction arises from electron-phonon coupling, according to Bardeen-Cooper-Schrieffer (BCS) theory. The space-dependent mean-field BCS Hamiltonian was developed by P. W. Anderson to study the interplay of superconductivity and inhomogeneities, where the pairing potential can be calculated via a self-consistent gap equation[14, 33]. Following this framework, we numerically simulate the effects of nonmagnetic disorder on superconductivity in the strong disorder limit by calculating the value of the pairing potential using a self-consistent gap equation.

To simulate the effects of nonmagnetic disorder on the 3D s-wave superconductor, we consider the negative-U Hubbard Hamiltonian[32] given by:

$$H = \sum_i (\varepsilon_i - \mu)\psi_i^\dagger \psi_i + t(\psi_i^\dagger \psi_{i+1} + \psi_{i+1}^\dagger \psi_i) - U\psi_{i\uparrow}^\dagger \psi_{i\downarrow}^\dagger \psi_{i\uparrow} \psi_{i\downarrow}$$

where $\psi_{i\sigma}^\dagger (\psi_{i\sigma})$ creates (annihilates) in the electron with spin σ on the $i$th site. $\mu$ denotes the chemical potential, $\varepsilon_i = \varepsilon_0 + V_i$ is onsite potential, $t$ represents the nearest neighbor hopping, and $U$ is the attractive interaction strength. Here $\varepsilon_0$ is a constant and the nonmagnetic disorder $V_i$ randomly distributed between $[-\frac{W}{2}, \frac{W}{2}]$ with $W$ is the disorder strength. By using the modified BCS mean-field theory, the mean field Hamiltonian becomes[33]

$$H_{mean} = \sum_{n\sigma} \varepsilon_n c_{n\sigma}^\dagger c_{n\sigma} + \sum_n \Delta_n c_{n\uparrow}^\dagger c_{\bar{n}\downarrow}^\dagger + h.c.$$

where the superconductor paring potential $\Delta_m = -U\langle c_{m\uparrow} c_{\bar{m}\downarrow}\rangle$ and the interaction matrix $I_{nm} = \sum_i \phi_{m\uparrow}^*(r_i)\phi_{\bar{m}\downarrow}^*(r_i)\phi_{n\uparrow}(r_i)\phi_{\bar{n}\downarrow}(r_i)$. The creation operators $c_{n\uparrow}^\dagger$ and $c_{\bar{n}\downarrow}^\dagger$ are responsible for creating electrons in the time-reversal partner eigenstates $\phi_{n\uparrow}$ and $\phi_{\bar{n}\downarrow}$ associated with the eigenvalue $\varepsilon_n$ for the non-interacting Hamiltonian. Then, the zero-temperature self-consistent gap equation is given by

$$\Delta_m = \frac{U}{2} \sum_{|\varepsilon_n|<\varepsilon_D} \frac{1}{\sqrt{\varepsilon_n^2 + \Delta_n^2}} I_{nm}\Delta_n,$$

and the local paring potential $\Delta_i = \frac{U}{2}\sum_n \frac{\Delta_n}{\sqrt{\varepsilon_n^2+\Delta_n^2}} \phi_{n\uparrow}(r_i)\phi_{\bar{n}\downarrow}(r_i)$. In our simulations, we set $t=1$, $U=2$, $\varepsilon_0=0$, $\mu=0$ and $\varepsilon_D=0.2$.

**Acknowledgements**

This work was supported by the National Natural Science Foundation of China (Grant No. 52272265) and the National Key R&D Program of China (Grant No. 2018YFA0704300). T.P.Y. thanks the support from the National Key R&D Program of China (Grant No. 2021YFA1401800) and the Natural Science Foundation of China (Grant No. 52272267, 52250402). C.-Z. Chen is the support by the National Key R&D Program of China Grant (No. 2022YFA1403700), and the Natural Science Foundation of Jiangsu Province Grant (No. BK20230066). The authors thank the Analytical Instrumentation Center (# SPST-AIC10112914), SPST, ShanghaiTech University and the Analysis and Testing Center at Beijing Institute of Technology for assistance in facility support. The authors thank the staffs from BL15U1 at Shanghai Synchrotron Radiation Facility for assistance during data collection.

**Author contributions**

Y.Q. and T.P.Y. conceived the project. Y.Z, and L.X.Z. synthesized the samples with help from N.Y. Q.W., W.C., and C.L.. J.C., Q.Z., and L.G. performed the STEM measurements. Y.Z, and L.X.Z. performed high-pressure measurements with help from C.P., S.Z., L.L.Z., and M.Z.. J. D., C.-Z.C. and J.W. performed the theoretical calculations. Y.Z., L.X.Z., C.-Z.C, and Y.C. analyzed the data and plot the figures. T.P.Y., Y.Q., C.-Z.C., T.X.Y., L.X.Z and Y.Z. wrote the manuscript with the contributions of all authors.

**Competing interests**

The authors declare no competing interests.